# Electromagnetically Induced Transparency in all-dielectric metamaterials due to nullifying of multipole moments

ANAR K. OSPANOVA[1,2,*], ALINA KARABCHEVSKY[2], ALEXEY A. BASHARIN[1,3,4]

[1] *National University of Science and Technology (MISiS), Department of Theoretical Physics and Quantum Technologies, 119049 Moscow, Russia*
[2] *Electrooptical Engineering Unit and Ilse Katz Institute for Nanoscale Science & Technology,*
*Ben-Gurion University, Beer-Sheva 84105, Israel*
[3] *National University of Science and Technology (MISiS), The Laboratory of Superconducting metamaterials, 119049 Moscow, Russia*
[4] *Politecnico di Torino, Department of Electronic and Telecommunications, Torino 10129, Italy*
*\*Corresponding author: anar.k.ospanova@gmail.com*

**Here, we propose novel transparency effect in cylindrical all-dielectric metamaterials. We show that cancellation of multipole moments of the same kind lead to almost zero radiation losses due to the counter-directed multipolar moments in metamolecule. . Nullifying of multipoles, mainly dipoles and suppression of higher multipoles results in ideal transmission of incident wave through the designed metamaterial. The observed effect could pave the road to new generation of light-manipulating transparent metadevices such as filters, waveguides, cloaks and more.**

http://dx.doi.org/10.1364/OL.99.099999

Narrow transmission peak called "transparency window" in the optical spectral range [1], is one of the most promising effects in nanophotonics. This effect provides new field in electronic and optical applications. Namely, slowlight propagation and long pulse delays for the storage of optical data in matter, frequency selectivity for narrow-band filters, enhanced nonlinear effects, strong light-matter interaction in photonics [2]. For the first time, observed in quantum systems, soon this phenomenon imitated in classical objects. Experimentally classical Electromagnetically Induced Transparency (EIT) obtained in metamaterials in microwave frequency range [3]. At the same time, metamaterials are manmade materials exhibiting unnatural properties like negative refraction, cloaking, strong field localization and others [4-8]. Since metamaterials are free for geometrical modifications, they may be tuned to reach narrow band transparency window corresponding to high Q-factor. Thus, next challenge is to create suitable structures with proper interaction for EIT-like phenomena.

One can demonstrate "transparency window" by inducing overlapping of electric and magnetic multipoles in plasmonic and all-dielectric particles [9]. This phenomenon called Fano-resonance arising due to the interference between different parts of constituent metamolecules [10,11]. Another technique to observe the EIT in metamaterials is "trapping" incident electromagnetic wave and exciting destructive interference between the same multipoles in metamaterials [12,13]. In addition, anapole mode can be defined as the third principal method of transparency produced by destructive interference between electric and toroidal multipole moments of the same amplitudes and angular momentum [14-16].

All aforementioned systems possess low radiative losses and exhibit transparency due to destructive interference of the main two multipoles of the same order and suppressing other multipoles. However, in this paper we propose metamaterial transparency effect due to nullifying of main excited dipole moments leading to almost zero radiative losses in all-dielectric metamaterials.

The unit cell of proposed metamaterial consists of four identical subwavelength high-index dielectric cylinders forming rhombus. For the demonstration of well-pronounced effect in THz frequency range, we choose the cylinders made of $LiTaO_3$ with dielectric permittivity of 41.4. The height of each cylinder assumed to be infinitely elongated. Electromagnetic response of our system is characterized by the displacement currents induced in each cylinder by an incident electromagnetic wave. Displacement currents cause electromagnetic scattering that becomes resonant due to the accurately chosen radius and permittivity of cylinders and the polarization of incident electromagnetic wave. This resonant behavior corresponds to Mie resonance emerging on cylindrical all-dielectric particles. In our case, parallel-polarized electromagnetic wave excites resonant electric dipolar moment in each high-index dielectric cylinder as can be seen in Fig 1 [17,18]. It gives an opportunity to use dielectric metamaterials as unique structure for artificial magnetism, magnetic and toroidal dipolar excitation as well as anapole mode [18-24].

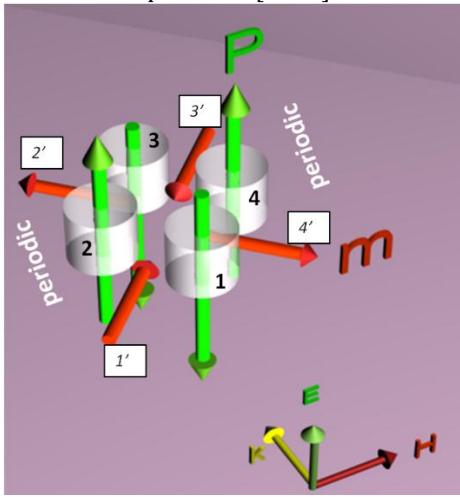

Fig.1. Illustration of the proposed metamolecule, consisting of 4 identical parallel dielectric cylinders. Electric component of incident wave polarized parallel to cylinder axis and induces counter-directed electric moments in each neighboring cylinder. Loops excite four magnetic dipolar moments that directed to and from the center of the metamolecule.

In our case, incident electromagnetic wave with E-field polarized parallel to cylinder axis induces counter-directed **P** electric moments in opposite [Fig 1] cylinders (1 and 3; 2 and 4) and forms two pair of displacement current loops in each metamolecule. Each pair of loops create two magnetic dipole (2' and 4') moments **m** directed out of center. Another pair of magnetic dipole moments emerges between cylinders from different pairs (1' and 3') and directed to the center of the unit cell. All magnetic moments of the studied system are orthogonal to cylinder axis. Each pairs of magnetic dipolar moments have opposite direction and eliminate each other. Therefore, the total magnetic response of the system disappears. In addition, the total electric dipolar response of metamolecule damped as well. This can be explained by the cancellation of displacement currents of neighboring cylinders from different current loops.

In previous works transparency was suggested as consequence of destructive interference between multipole moments as in [10, 11, 15]. However, our findings presented here show that transparent metamaterials can be designed without any dipolar response just due to the nullifying of each kind of dipoles. This provides zero radiation losses and such metamaterial becomes transparent in optical frequency range.

Our proposed system demonstrates strong electromagnetic interaction due to the near-field coupling of close located high-index dielectric cylinders. In THz frequency range $LiTaO_3$ distinguished by low level of dissipation losses. The unit cell parameters scaled to obtain pronounced dipolar response at the expense of higher order multipoles. Each cylinder has radius of 5 μm and center-to-center distance with neighboring within metamolecule cylinders of 12 μm. Period between clusters is 60 μm. Clusters are surrounded by air or vacuum medium. We assume indefinitely elongated height of cylinders allowing to consider two-dimensional structure. Electromagnetic properties of metamaterial are calculated by commercial Maxwell's equation solver HFSS using standard modeling approach, where the whole structure described by replicating unit cell properties using periodic boundary conditions.

Transmission spectrum of our metamaterial is depicted on Fig 2. The sharp narrow transmission peak at 2.2446 THz corresponds to the emergence of the transparency effect. This resonance has amplitude of 1 and width of 0.0017 THz, which corresponds to very high Q-factor value of 1320.

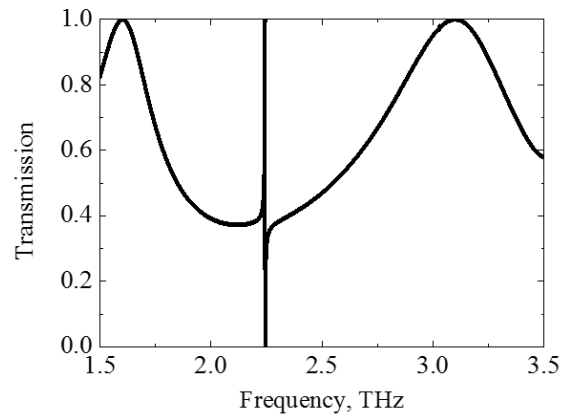

Fig. 2. Transmission spectrum for metamolecule in THz frequency range. Narrow sharp transparency peak corresponds to f=2.2446 THz with Q=1320.

Field map at this value shows opposite distributed electric field strengths in every neighboring cylinder indicating on occurrence of displacement currents loops in each pair of cylinders [Fig 3a]. Direction of field strength of cylinder 1 coincides with the direction in the cylinder 3, while both 2 and 4 cylinders have opposite field distribution directions. It means that these cylinders, let us say pair 1 and 2, 2 and 3, 3 and 4, 4 and 1, forms four loops of displacement currents in every unit cell of the system. In turn, two loops generate two magnetic dipolar responses directed out of center of the unit cell (2' and 4' of Fig.3b). Other two loops of displacement currents generate other pair of magnetic dipole moments that directed toward center (1' and 3' of Fig.3b). Field map on Fig. 3b shows magnetic field distribution in the unit cell. The strong magnetic field localization between cylinders indicates the near field coupling between them. These oppositely directed magnetic dipole moments have zero contribution in metamolecule dipolar response. There is also zero total electric dipolar response, since electric field strength of cylinders 1 and 3 cancels electric field strength of cylinders 2 and 4.

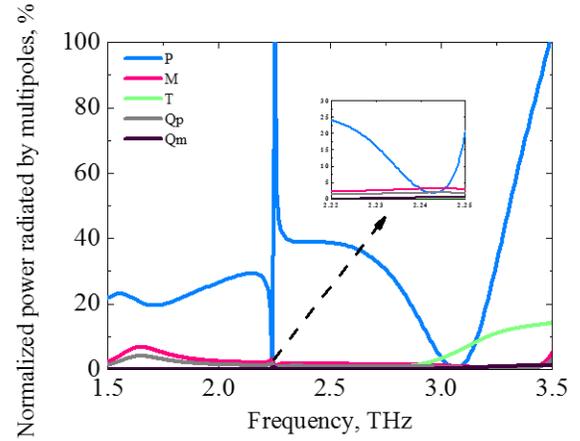

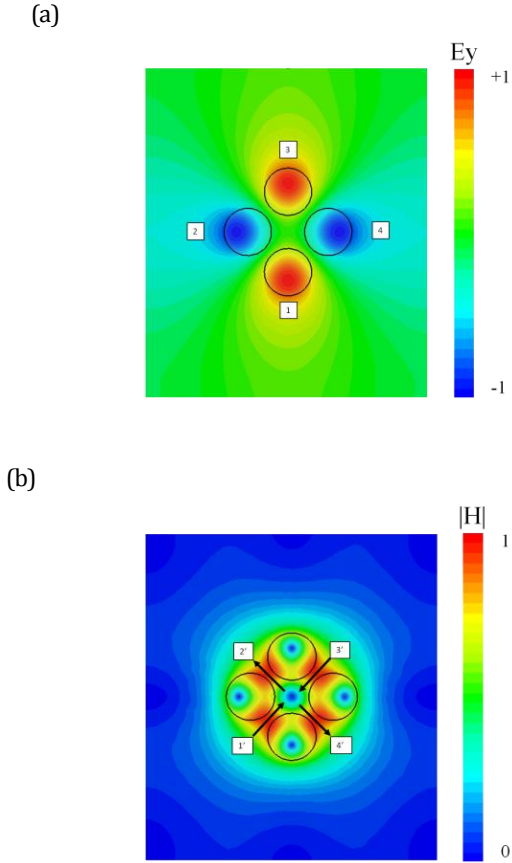

Fig. 3 Field maps of (a) *y*-component (along cylinders axis) of electric field and (b) absolute value of magnetic field intensities.

To confirm our assumption on nullifying the same kind multipoles we carried out multipolar decomposition up to second order of multipoles radiated by metamolecule at resonant frequency value. Figure 4 shows normalized power of near-field distribution of metamolecule up to second order multipoles. Resonant frequency corresponds to the second peak of transmission spectrum. At the resonant frequency f=2.2446 THz one can see narrow dip of electric dipole moment accompanying with suppression of quadrupole moments and almost zero value of magnetic and toroidal dipoles. Accordingly, in this frequency range radiation losses tends to zero and this leads to full transparency at f=2.2446 THz. This confirms our assumption that transparency of metamaterial can be achieved due to nullifying of all multipoles contemporaneously. This effect accompanied by very high Q-factor, corresponding to narrow transparency window [Fig. 2]. We note, that high Q-factor value has been previously considered for all-dielectric system with broken symmetry in optical range. In ref. [25] the symmetry of cubic dielectric metasurface was broken and placed on substrate to obtain resonance between two modes. As the result, very high-Q Fano resonance of – 600 was obtained. Our metamaterial distinguished with avoidance of such sophisticated manufacturing by placing cylinders in defined sequence. This elegant approach is obviously technologically simple.

Fig. 4. Normalized power of near-field distribution of metamolecule up to second order multipoles. Sharp dip of electric dipole moment (light blue curve) corresponds to transparency peak at 2.2446 THz. There is also reduction of electric (grey curve) and magnetic (violet curve) quadrupole moments and magnetic (pink curve) and toroidal (green curve) dipolar moments. Inset shows excited multipoles close to f=2.2446 THz.

Moreover, the behavior of third peak which is close to f=3 THz gives the following outcomes: from 3 THz to 3.14 THz frequency range, which coincides to third peak of transmission spectra [Fig. 2, 4], it becomes higher among other multipoles leading to the establishing of toroidal mode.

In this work, we presented new transparency effect arising from extinguishing of all kinds of multipoles at the same time. To confirm our assumption, we proposed metamaterial structure and carry out multipolar decomposition. Simulation results show that at transparency peak of Q=1320 there is suppression of all type of multipole moments up to zero pretending non-radiative system. These results are extremely important in designing invisible to external observer systems that could be used in many applicable fields as nanophotonics, plasmonic and applied optics.

ACKNOWLEDGMENT. This work was supported by the Ministry for Education and Science of the Russian Federation, in the framework of the Increase Competitiveness Program of the National University of Science and Technology MISiS under contract numbers K4-2015-031, the Russian Foundation for Basic Research (Grant Agreements No. 16-32-50139 and No. 16-02-00789) and partially by the start-up grant of A.K. In addition, this work has been partially supported by the Joint Projects for the internationalization of Research launched by the Politecnico di Torino with the financial support of the Compagnia di San Paolo, project title: "Advanced Non-radiating Architectures Scattering Tenuously And Sustaining Invisible Anapoles (ANASTASIA)". The work on the multipoles decomposition investigation of the metamolecules was supported by Russian Science Foundation (project 17-19-01786). The results leading to this manuscript were obtained during the joint PhD program between BGU (under supervision of A.K.) and NUST MISiS (under supervision of A.A.B).

**References**

1. Stephen E. Harris, Phys. Today 50, 7, 36 (1997).
2. S. E. Harris, J. E. Field, and A. Imamoglu, Phys. Rev. Lett. 64, 1107 (1990).
3. N. Papasimakis, V. A. Fedotov, N. I. Zheludev, and S. L. Prosvirnin, Phys. Rev. Lett. 101, 253903 (2008)
4. Y. Galutin, E. Falek and A. Karabchevsky, Invisibility Cloaking Scheme by Evanescent Fields Distortion on Composite Plasmonic Waveguides with Si Nano-Spacer, Sci. Rep., accepted (2017)
5. D. R. Smith,1 J. B. Pendry,2 M. C. K. Wiltshire, Science, 305(5685), 788-92 (2004)
6. Yongmin Liua  and  Xiang Zhang, Chem. Soc. Rev., 40, 2494-2507 (2011)
7. J.B. Pendry, Phys Rev Lett, 85, 3966, (2000)
8. C. M. Soukoulis and M. Wegener, Nat. Photonics 5, 523 (2011).
9. Boris Luk'yanchuk, Nikolay I Zheludev, Stefan A Maier, Naomi J Halas, Peter Nordlander, Harald Giessen, Chong Tow Chong, Nature materials, 9, 9, 707 (2010)
10. Shuang Zhang, Dentcho A. Genov, Yuan Wang, Ming Liu, and Xiang Zhang, PRL 101, 047401 (2008)
11. P Tassin, L Zhang, R Zhao, A Jain, T Koschny, CM Soukoulis, Phys. Rev. Lett. 109 (18), 187401 (2012)
12. Fedotov VA1, Rose M, Prosvirnin SL, Papasimakis N, Zheludev NI., Phys. Rev. Lett. 99(14), 147401 (2007)
13. Nikitas Papasimakis and Nikolay I. Zheludev, Optics and Photonics News 20, 10, 22-27 (2009)
14. N. Papasimakis, V. A. Fedotov, V. Savinov, T. A. Raybould & N. I. Zheludev, Nature Materials 15, 263–271 (2016)
15. V. A. Fedotov, A. V. Rogacheva, V. Savinov, D. P. Tsai and N. I. Zheludev, Sci. Rep. 3, 2967 (2013).
16. Nikita A. Nemkov, Alexey A. Basharin, and Vassili A. Fedotov, Phys. Rev. B 95, 165134 (2017)
17. C. F. Bohren and D. R. Huffman, (Wiley-Interscience, New York, 1983).
18. Yuri Kivshar and Andrey Miroshnichenko, Philos Trans A Math Phys Eng Sci , (2017)
19. A.A. Basharin, I.V. Stenischev, Toroidal response in all-dielectric metamaterials based on water, Sci. Rep., accepted (2017)
20. Alexey A. Basharin, Maria Kafesaki, Eleftherios N. Economou, Costas M. Soukoulis, Vassili A. Fedotov, Vassili Savinov and Nikolay I. Zheludev, Phys. Rev. X 5, 011036 (2015).
21. W Liu and Yuri S.Kivshar, Philos Trans A Math Phys Eng Sci. 375, 2090, (2017)
22. Boris Luk'yanchuk, Ramón Paniagua-Domínguez, Arseniy I. Kuznetsov, Andrey E. Miroshnichenko, Yuri S. Kivshar, Philos Trans A Math Phys Eng Sci.  375, 2090, (2017)
23. Wei Liu, Andrey E. Miroshnichenko, arXiv:1704.06049 [physics.optics]
24. Andrey E. Miroshnichenko, Andrey B. Evlyukhin, Ye Feng Yu, Reuben M. Bakker, Arkadi Chipouline, Arseniy I. Kuznetsov, Boris Luk'yanchuk, Boris N. Chichkov, and Yuri S. Kivshar, Nat Commun.6, 8069, (2015)
25. Salvatore Campione, Sheng Liu, Lorena I. Basilio, Larry K. Warne, William L. Langston, Ting S. Luk, Joel R. Wendt, John L. Reno, Gordon A. Keeler, Igal Brener, and Michael B. Sinclair, ACS Photonics, 3 (12), 2362–2367 (2016)